\documentclass[twocolumn,twoside,letterpaper,11pt]{article}
\usepackage{graphicx,xrrc,natbib}
\usepackage{times}

\begin{document}


\title{CEN A AND ITS INTERACTION WITH THE X-RAY EMITTING INTERSTELLAR
MEDIUM}


\author{D.M. Worrall} 
\institute{Dept.~of Physics, University of Bristol} 
\address{Tyndall Avenue, Bristol BS8 1TL, U.K.} 
\email{d.worrall@bris.ac.uk} 

\author{R.P. Kraft, M. Birkinshaw, M.J. Hardcastle, 
        W.R. Forman, C. Jones, S.S. Murray}
\email{rkraft@cfa.harvard.edu, mark.birkinshaw@bris.ac.uk, 
           m.hardcastle@bris.ac.uk,}
\email{wforman@cfa.harvard.edu,
        cjones@cfa.harvard.edu, ssm@cfa.harvard.edu}


\maketitle

\abstract{ As the closest radio galaxy, Centaurus A is a powerful
laboratory for the X-ray study of radio-emitting structures and their
interactions with the hot interstellar medium (ISM).  This paper
details our interpretation of the remarkable X-ray enhancement which
caps the inner southwest radio lobe, at a radius of about 6~kpc from
the galaxy center.  The shell of X-ray-emitting gas is hotter than the
ambient ISM, and overpressured by a factor of 100.  We argue that it
is heated compressed material behind the supersonically-advancing bow
shock of the radio lobe, the first example of the phenomenon to be
clearly detected.  The results demonstrate that Cen A is actively
re-heating nearby X-ray-emitting gas.  The shell's kinetic energy is
$\sim 5$ times its thermal energy, and exceeds the thermal energy of
the ISM within 15~kpc of the center of the galaxy.  As the shell
dissipates it will have a major effect on Cen~A's ISM, providing
distributed heating.  }

\section{Introduction}

There is much current interest in the possibility that radio sources
heat the interstellar and intercluster medium.  Such heating would
help to explain the weakness or absence of lines from gas cooling
below 1~keV in the densest central regions of galaxies and clusters
\citep[e.g.,][]{peter01}.  A good place for heating would be behind
the bow shock of a supersonically expanding radio structure.  In this
paper we outline the simple theory and its application to Cen~A, the
first source to show clear and direct X-ray evidence of supersonic
expansion and heating.

\section{Theory}

\begin{figure}
  \begin{center}
    \includegraphics[width=\columnwidth]{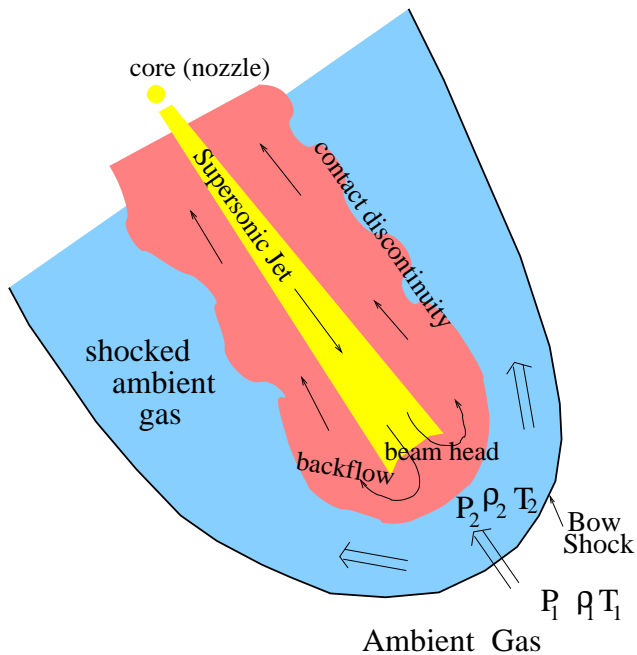}
    \caption{\small In the standard model for powerful radio sources,
    a supersonic jet (yellow) terminates at the beam head, producing a
    radio hotspot.
    Provided the shocked radio-emitting fluid forming the radio lobe
    (pink) has enough internal energy or momentum density to drive
    a leading bow shock, ambient X-ray-emitting gas will be heated as
    it crosses the shock to fill the blue region. }
    \label{fig:supersonic}
  \end{center}
\end{figure}

The standard model for the expansion of a powerful radio source
powered by a jet which is supersonic with respect to the
X-ray-emitting interstellar medium (ISM) is illustrated in
Fig.~\ref{fig:supersonic}.  The jet terminates at the beam head (in a
feature identified as the radio hotspot) where the jet fluid passes
through a strong shock to inflate a cocoon of radio-emitting plasma.
The energy and momentum flux in the flow is normally expected to be
sufficient to drive a bow shock into the ambient medium ahead of the
jet termination shock.  In the rest frame of the bow shock, ambient
gas is heated as it crosses the shock to fill a region surrounding the
lobe of radio-emitting plasma.

The sound speed in gas of temperature $T$ is given by

$$c_{\rm s} = \sqrt{\gamma kT \over \mu m_{\rm H}}\eqno(1)$$

where $\gamma$ is the ratio of specific heats, which for a monatomic
non-relativistic gas is 5/3, $k$ is the Boltzmann constant, $m_{\rm
H}$ is the mass of the hydrogen atom, and $\mu m_{\rm H}$ is the mass
per particle, which is 0.6 $m_{\rm H}$ for a gas with normal cosmic
abundances.  Under these conditions, $c_{\rm s} \approx 516 
({kT/{\rm keV}})^{1/2}$ km s$^{-1} \approx 0.54 ({kT/{\rm
keV}})^{1/2}$ kpc Myr$^{-1}$.

The Mach number of the speed of advance, $v_{\rm adv}$, of the bow
shock into the ambient medium is ${\cal M} = v_{\rm adv} / c_{\rm s}$,
which in convenient units can be expressed as

$${\cal M} \approx  580 (v_{\rm adv}/c) (kT/{\rm keV})^{-1/2}\eqno(2)$$

where $c$ is the speed of light.  In a simple application of the
Rankine-Hugoniot conditions for a strong shock
\citep[e.g.,][]{spitzer78}, the pressure, density, and temperature,
respectively, in the unshocked (subscript 1) and shocked (subscript 2)
regions at the head of the bow shock are related by

$$P_2/P_1 = (5 {\cal M}^2 -1)/4 \eqno(3)$$

$$\rho_2/\rho_1 = 4 {\cal M}^2/ ({\cal M}^2 +3) \eqno(4)$$

$$T_2/T_1 = (5 {\cal M}^2 -1)({\cal M}^2 +3)/16 {\cal M}^2 \eqno(5)$$

for a monatomic gas.  

If the density in an X-ray-emitting gas is
described by the proton number density, $\eta_{\rm p}$, then the
thermal pressure is given by

$$P = \eta_{\rm p} kT /X \mu \eqno(6)$$

where $X$ is the abundance of hydrogen by mass, which is 0.74 for
normal cosmic abundances \citep[e.g.,][]{birk93}.

For hot X-ray emitting gas where the cooling in line radiation is
unimportant, the X-ray emissivity, $\cal E$, between energies $E_1$
and $E_2$ depends on temperature and proton density approximately as

$${\cal E} \propto \eta_{\rm p}^2 T^{0.5} (e^{-E_1/kT} -
e^{-E_2/kT})\eqno(7)$$

where the weak energy dependence of the
Gaunt factor is ignored.  At temperatures below $\sim 1$~keV, line
radiation cannot be ignored, and the plasma models incorporated
into X-ray-spectral-fitting programs such as {\sc XSPEC} can be
used to find the dependence of the emissivity on energy.
We find that in the energy band 0.8--2 keV, where
${\it Chandra\/}$ and XMM-Newton are most sensitive, for
an ${\cal M} = 4$ shock, the X-ray emissivity contrast between shocked
and unshocked gas is a factor of 3 higher if the ambient gas is at
0.29 keV (as found for Cen~A, Table~\ref{tab:physical}) than if the
external medium has a typical cluster temperature of 4~keV.

Complications apply in reality, and in practice these are difficult to
treat even with data from observatories as powerful as {\it Chandra\/}
and XMM-Newton.  Firstly, there is observational evidence that in
supernova remnants with shocks of comparable Mach number to that found
for Cen~A the post-shock electrons are cooler than the ions
\citep[e.g.,][]{hwang02, rakowski03}. This is not taken into into
account in our modelling. Secondly, the simple Rankine-Hugoniot
equations that we quote and apply do not take into account the fact
that the bow shock around a lobe is oblique away from its head, with a
consequent change in the jump conditions and the emissivity contrast
\citep[e.g.,][]{williams91}.  However, if Cen A's shell represents a
spherical expansion rather than a lobe structure, then the
shock should be normal everywhere and the equations above will hold.

\section{Historical Perspective}

It is possible to interpret ROSAT-detected X-ray cavities coincident
with the inner parts of the radio lobes of Cygnus~A as due to an
emissivity contrast between bow-shock heated gas outside the lobes
(heated to temperatures above the ROSAT energy band) and the more
easily detected ambient cluster medium \citep{car94}. However, the
parameters of the shock are not effectively constrained by the ROSAT
X-ray data.  Similarly, {\it Chandra\/} observations of Cygnus~A find
gas at the sides of the lobes to have $kT \sim 6$~keV, slightly hotter
than the value of 5~keV from ambient medium at the same cluster
radius, possibly indicating cooling after bow-shock heating, but again
the data do not usefully constrain model parameters \citep{smith02}.
Other reports of lifting of gas (leading to eventual heating) by radio
lobes or hot bubbles are not thought to involve supersonic expansion
\citep[e.g.,][]{chur01, quilis01}, although it has been suggested
recently that filaments of hot gas in the atmosphere of M~87 are
slowing from supersonic speeds after ejection from the galaxy center
\citep{forman04, kraft04}. 

The first and best example of a shell of
heated gas which can reasonably be attributed to supersonic expansion
is in Cen~A.  High-quality {\it Chandra\/} and XMM-Newton data
\citep{kraft03, kraft04} provide the temperature and density
constraints needed to test the model and measure the supersonic
advance speed of the bow shock responsible for the heating.

\section{Cen~A}

Cen~A is our nearest radio galaxy, at a distance of 3.4~Mpc
\citep{israel98} so that
1~arcsec corresponds to $\sim 17$~pc, and is an example
of a low-power radio galaxy.  In such sources the
radio jet is normally expected to have slowed considerably through
entrainment of ambient material \citep{bickn94}, at which point the
model described by Fig.~\ref{fig:supersonic} no longer holds.
The full extent of Cen~A's radio
emission covers several degrees on the sky \citep{junkes93}.
Within this lies a sub-galaxy-sized double-lobed inner structure
\citep{burns83} with a predominantly one-sided jet to the northeast
and a weak counter-jet to the southwest \citep{hard03}, embedded in a
radio lobe with pressure $1.4 \times 10^{-12}$~Pa or more, greater
than the pressure in the ambient ISM ($\sim 1.8 \times 10^{-13}$~Pa;
Table~\ref{tab:physical}), and so which should be surrounded by a
shock.  Around this southwest lobe there is a shell of X-ray emitting
gas which appears to have the geometry of the shocked ambient gas in
Fig.~\ref{fig:supersonic} \citep{kraft03, kraft04}.  Although the
capped lobe is around the weak counterjet, so it is not evident that
the lobe is being thrust forward supersonically with respect to the
external interstellar medium (ISM) by the momentum flux of an active jet, 
the current high internal pressure in the radio lobe
ensures its strong expansion.

\begin{table*}[t]
  \begin{center}
  \caption{Physical parameters of the gas in various regions of Cen A}\medskip
  \label{tab:physical}
  \begin{tabular}{lcclc}
Structure & $kT$ & $\eta_{\rm p}$ & Pressure &
  0.4--2 keV  \\
        & (keV) & Proton density & (Pa)$\dag$ &
   relative \\
&&(m$^{-3}$) &&emissivity, $\cal E$ \\
    \hline\hline
ISM (measured) & 0.29 & 1700 & $1.8 \times 10^{-13} \ddag$ (thermal) &
  1 \\
Behind bow shock (inferred) & 6.8 & 6530 & 
$2.1 \times 10^{-11}$ (thermal+ram $\ast$) & 13\\
Shell (measured) & 2.9 & 20000 & $2.1 \times 10^{-11}$ (thermal) & 127\\
    \hline
$\dag$  1 Pascal = 10 dyn cm$^{-2}$ & \multicolumn3{c}{$\ddag$ incorrectly
  reported in Table~5 of \citet{kraft03}}&$\ast$ ~~$\rho_1 v^2_{\rm
  adv} / 4$\\
\multicolumn4{l}{
By comparison, the minimum-energy pressure in the radio lobe is $\sim
  1.4 \times 10^{-12}$ Pa.}\\
  \end{tabular}
  \end{center}
\end{table*}

\section{Application of the model to Cen~A}

The temperature, proton density and pressure of the ambient ISM and
the X-ray shell, taken from \citet{kraft03} are given in
Table~\ref{tab:physical}.  The ambient medium is measured to have
$\eta_{\rm p}\sim 1.7 \times 10^3$ m$^{-3}$ and $kT = 0.29$ keV,
whereas the shell is ten times hotter, at $kT = 2.9$ keV, and
twelve times denser, with $\eta_{\rm p}\sim 2 \times 10^4$ m$^{-3}$.
From equations (4) and (5) above, we see that temperature and density
measurements for both the ambient medium and the shocked gas directly
test shock heating, since only two of the four parameters are required
to measure the Mach number, and the other two test the model.  

The most straightforward application of the equations finds an
inconsistency, since the densities and temperatures are not
self-consistent.  The shell's density and temperature are wrong for
gas directly in contact with the bow shock.  However, we can find a
Mach number consistent with shocking the gas to a temperature and
density such that the combined thermal and ram pressure is in pressure
equilibrium with the thermal pressure of the detected shell: ${\cal M}
= 8.5, v_{\rm adv} \approx 2400$ km s$^{-1}$.  The post-shock
temperature is $kT_2 \sim 6.8$~keV.  The 6.8~keV gas flows back from
the shock, into the X-ray-detected shell at 2.9~keV.  The
characteristics of this undetected hotter gas are given in
Table~\ref{tab:physical}.  In this table we also quote estimates of
the relative X-ray emissivity (per unit volume) of gas in the
different structures over the 0.4-2 keV energy band, where the {\it
Chandra\/} response peaks and is relatively flat.  The gas directly
behind the bow shock has a predicted emissivity that is an order of
magnitude fainter than that in the shell, accounting for its absence
in our measurements.

\section{Conclusions}

\begin{itemize}

\item A hot shell of X-ray-emitting gas caps the southwest radio lobe of
Cen~A.

\item Shock jump conditions for an advance speed of $\sim 0.008$c 
(${\cal M} = 8.5$)
are satisfied if the shell is 
in pressure balance with unseen gas at $kT \sim 6.8$~keV behind the
bow shock.  The emissivity of the 6.8 keV gas is too low to separate
its X-rays from those of the ten-times-brighter gas of the shell.

\item The radiative timescale for material in the shell ($\sim 2
\times 10^9$
yrs, using equation 5.23 of \citet{saraz86}) is large compared with
the lobe expansion time ($< 2.4 \times 10^6$
years), so the material in the shell behaves as an adiabatic
gas \citep[e.g.,][]{alex02}.

\item The cooling has improved the shell--ISM contrast
by an order of magnitude, an important factor in the detection and
modelling of the shell.

\item The lobe is (or was) powered by energy deposition from a jet,
and is overpressured relative to the ambient ISM.  Monitoring the
intensity and searching for proper motions in the knots of the weak
counterjet on this side of the source would provide important
diagnostics for establishing the current state of the jet within the
southwest lobe.

\item To order of magnitude, the mass in the hot shell is consistent
with material swept up from the ISM.  There must have been sufficient
time since any earlier epoch of lobe expansion in the region for the
ISM to be replenished.

\item The shell is overpressured compared with the minimum-energy
pressure in the radio lobe (in magnetic field and radiating electrons)
by a factor of $\sim 10$.  There is no particular reason to think that
a dynamical object, seen in a snapshot, should be in a state of
minimum energy.  However, if we do assume minimum energy in the lobe,
and that the shell has reached equilibrium [but note that
the sound-crossing time in the shell (thickness $\sim 0.3$
kpc, $c_{\rm} \sim  9
\times 10^{-7}$ kpc yr$^{-1}$)
is about 15 per cent of the maximum time we estimate it has taken the lobe to
reach its current size], the shell's  overpressure relative to the radio
lobe could be balanced by the ram pressure from internal motions in
the lobe for a moderate relativistic proton loading.

\item The shell's kinetic energy is $\sim 5$ times its thermal
energy, and exceeds the thermal energy of the ISM within 15 kpc of the
centre of the galaxy.  As the shell dissipates, most of the kinetic
energy should ultimately be converted into heat and this will
have a major effect on Cen~A's ISM, providing distributed heating.

\end{itemize}

\section*{Acknowledgments}

DMW is grateful to Martin Laming for pointing out the
possible inaccuracies of assuming that the post-shock electron and ion
temperatures are identical.
She thanks the organizers for supporting her accommodation and
subsistence during
the workshop.  MJH thanks the Royal Society for a research fellowship.

\end{document}